\documentclass[twocolumn,,floatfix,8pt]{revtex4}
\usepackage{graphicx}
\usepackage{amssymb}
\usepackage{epstopdf}
\usepackage{float}
\usepackage{subfigure}
\begin{document}

\title{Experimental study of the impact of historical information in human coordination}
\author{Manuel Cebrian}
\author{Ramamohan Paturi}
\author{Daniel Ricketts}
\affiliation{Department of Computer Science and Engineering, University of California, San Diego, La Jolla, CA 92093}

\begin{abstract}
We perform laboratory experiments to elucidate the role of historical information in games involving human coordination. Our approach follows prior work studying human network coordination using the task of graph coloring. We first motivate this research by showing empirical evidence that the resolution of coloring conflicts is dependent upon the recent local history of that conflict. We also conduct two tailored experiments to manipulate the game history that can be used by humans in order to determine (i) whether humans use historical information, and (ii) whether they use it effectively. In the first variant, during the course of each coloring task, the network positions of the subjects were periodically swapped while maintaining the global coloring state of the network.
In the second variant, participants completed a series of 2-coloring tasks, some of which were restarts from checkpoints of previous tasks. Thus, the participants restarted the coloring task from a point in the middle of a previous task without knowledge of the history that led to that point. We report on the game dynamics and average completion times for the diverse graph topologies used in the swap and restart experiments. 
\end{abstract}

\maketitle

\section{Introduction}
There are numerous examples in which humans must coordinate in order to accomplish a collective goal. National political parties coordinate through networks of local party chapters. The Obama campaign famously used Twitter and Facebook to coordinate with supporters and potential followers. Companies must choose which software to use, balancing the benefits of a useful application with the disadvantages of being incompatible with other companies. Friends perform decentralized communication to decide on which bar to frequent.

In many of these cases, full communication between parties is not possible. Instead, each person can only communicate with a small subset of the larger group, thus forming a communication network over which coordination must occur.

We would like to understand the decision making process that humans employ when coordinating over networks. Much of the prior modeling work has assumed that humans base their strategic decisions solely on the current local state of their network, e.g. on the current behavior of friends, family, or colleagues. But when making choices in life we also often take into account past behavior of those with whom we interact. Over time, we develop rich behavioral models of friends, family, companies, and political parties. These models of our network neighbors influence how we ourselves behave.

Thus it seems natural to try to understand how the historical behavior of network neighbors influences the ability to coordinate over a network. Do humans use historical information in coordination games, and do they use it to their advantage?

To begin answering these questions, we have conducted a series of human-subject experiments and simulations. We follow prior work by Kearns {\em et al.} \cite{Kearns11082006,Judd09082010,kearns2009behavioral}, and follow-ups \cite{mccubbins2009connected,enemark2011does},  in modeling human network coordination using he task of graph coloring. Our experiments attempt to manipulate the game history that can be used by humans in order to determine (i) whether humans use historical information, and (ii) whether they use it effectively.

When coordinating over a network, humans may employ several different uses of history. For example, over the course of many coordination tasks, they may develop models of coordination that influence their strategic choices. However, in this work we are interested in a different use of history. In particular, we are interested in quantifying how much {\em recent} history humans use to influence their decisions in the {\em current} coordination task.

Previous work has addressed history usage in human coordination tasks over networks. Israeli et al. \cite{israeli2010low} showed that simple algorithms that could plausibly be executed by humans benefit from small amounts of history. Other work focuses on generating predictive models of human behavior~\cite{Duong:2010:HGM:1838206.1838364,duonglearning} or determining whether participants provided with algorithmic instructions achieve significantly better performance \cite{mao2011human}. Our work focuses on detecting and understanding the effect of history usage.


\begin{figure*}[!t]
\includegraphics[width=0.55\textwidth]{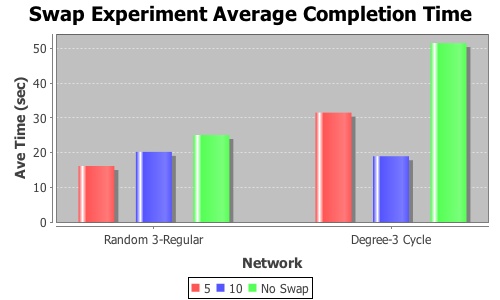}
  \caption{The averages completion time of the {\em swap} experiment for each network topology.}
  \label{AveSwp}
\end{figure*}

\section{Evidence for history usage in human coordination}
\label{confbias}

In each of the human-subject experiments we conducted, each of 16 subjects sat at a computer through which he could control the color of a node in a network. Each subject was only able to see the colors of his immediate network neighbors, so he could not explicitly know the global coloring state of the network. All subjects received one dollar if they were able to 2-color the network in under three minutes. No subjects received any money if this did not occur. During each three minute session, subjects could change their own color as often as they wanted while seeing, in real time, their neighbors' color changes. The sessions were repeated a number of times with the same set of participants over the course of two hours. We call this the standard protocol.

\begin{figure*}[ht]
\centering
\subfigure[Swap experiment topologies (from left to right): random 3-regular, cycle with cords.]{
\includegraphics[width=0.48\textwidth]{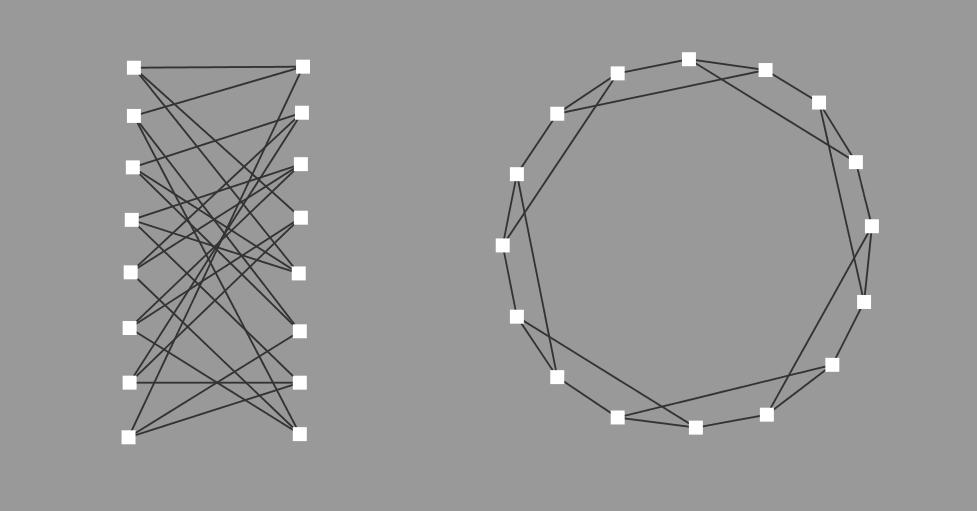}
\label{SwapTopos}
}
\subfigure[Restart experiment topologies (from left to right): line, barbell, simple cycle.]{
\includegraphics[width=0.48\textwidth]{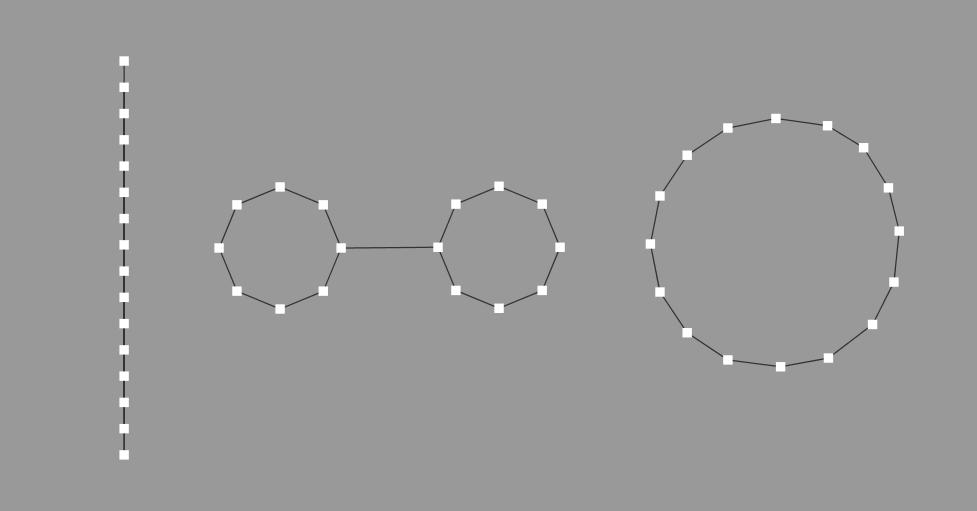}
\label{RestartTopos}
}
\label{fig:topos}
\caption{Experimental topologies.}
\end{figure*}

We have conducted two such human-subject 2-coloring experiments. We used five different network topologies over the two experiments. In the first experiment, subjects performed the 2-coloring task on a random 3-regular bipartite graph and on a simple cycle in which each node had on extra cord (figure \ref{SwapTopos}). In the second experiment, subjects performed the tasks on a simple cycle with no cords, a barbell network consisting of two cycles connected by a single edge, and a line graph (figure \ref{RestartTopos}).

%

In the two experiments, some of the sessions had a protocol that was slightly modified from the standard protocol. The modified protocols were tailored to understand usage of history by the subjects. These modified protocols are described in the following section. First, we provide a simple analysis of the sessions with the standard protocol. This analysis gives evidence that subjects use historical information when resolving conflicting edges, showing a strong directionality to the traversal of conflicts through the networks.

A conflicting edge is one whose two end points have the same color. The analysis considers each time a conflicting edge is resolved, i.e. when one of the endpoints changes color so that the endpoints are no longer the same color. If we assume that subjects do not use any history of past events, then by symmetry each end point is equally likely to resolve the conflict by changing its own color.

However, it is conceivable that subjects use history of a conflicting neighbor's most recent color change when making the decision to resolve the conflict. Indeed, in post-experimental questionnaires, many subjects did report using neighbors' most recent color change as a factor in their own color choices.

This leads to a simple analysis of the data, one which allows the timing of a neighbor's most recent color change to be a factor in a subject's resolution of conflicts. We count the number of times that the following occurs: the endpoint that resolves a conflicting edge has switched his color less recently than the other endpoint of the conflicting edge. Then divide this by the total number of times a conflicting edge is resolved. We call this fraction the {\em conflict resolution bias} Again by symmetry, if subjects do not use any history, the conflict resolution bias should be $0.5$. Subjects would not be concerned with how recently their neighbors have changed.

We can compute the conflict resolution bias for the standard protocol sessions for each of the two experiments that we have run. Each experiment used a different set of 16 subjects. For one experiment the conflict resolution bias was 0.77 over 930 instances of conflicting edges being resolved. For the other experiment, the conflict resolution bias was 0.41 over 523 instances of conflicting edges being resolved. Both values are different than the bias of 0.5 that is expected if subjects use no history. This suggests that subjects are using recent history. It also implies that conflicts travel in the network with a certain directionality.

However, the values are also on opposite sides of 0.5. This suggests the following explanation. Subjects use the timing of their neighbors' recent moves when making decisions, but they use this timing information in different ways. Some subjects may prioritize resolving conflicts with stably colored neighbors while other subjects resolve conflicts with neighbors who have changed recently. These two strategies have opposing effects on the conflict resolution bias. It may be the case that the distribution of the two strategies varies from group to group and thus so does the conflict resolution bias.

This analysis only captures a very simple use of history. We hope that the modified experiments presented in section \ref{experiments} will detect and show the effect of potentially more elaborate uses of history.


\section{Experiments}
\label{experiments}

We have conducted two variants of these graph coloring experiments in order to understand the use of history in human coordination games. In the first variant, which we call the {\em swap experiment}, the network positions of the subjects were periodically swapped during the three minute sessions, while maintaining the global coloring state of the network. In other words, every $k$ seconds, the subjects were randomly assigned to a node in the network. The color of each node in the network remained constant across the swap. Thus, a subject may control a red node prior to a swap and a blue node after the swap though he made no decision to change color. With high probability each subject's neighbors changed after each random assignment, so any history of neighbor behavior is effectively erased. Thus, subjects could only use historical knowledge from a limited time window since the last swap. This window was at most $k$ seconds.

We varied the frequency with which subjects are swapped in order to test the effect of changing the size of the window of historical information that can be used by the subjects. Within each session, the swapping frequency remained constant, but different sessions had different swapping frequencies. In the single experiment that we have run, we swapped subjects every $5$ seconds, every $10$ seconds, and never.

The instances of swapping can be potentially confusing for the subjects, so we provided an interface that allowed them to have a "context switching" period. When the swap occured, each subject was shown a screen that displayed his new node's color and each of his new neighboring node's colors. This screen was grayed out for a period of $3$ seconds during which time the subjects could not change their colors. After the $3$ second context switching period, the game resumed. The $3$ second context switching time did not count towards the three minute session time limit.

We used two different topologies (figure \ref{SwapTopos}) for the swap experiment, a random 3-regular graph and a cycle with one cord per node. These topologies were designed with two goals in mind. High diameter graphs tend to have longer expected completion time for 2-coloring \cite{Kearns11082006}. This is advantageous for the swap experiment because it maximizes the number of swaps that will occur before completion, thus maximizing the number of treatments per session. We used constant degree graphs in order to reduce confusion and unintended treatments from swapping subjects between nodes of varying degrees.

In the second variant, the experiment proceeded in two phases. In the first phase, the subjects performed a series of 2-coloring sessions in which all network nodes begin with no color. This is the standard protocol described in section \ref{confbias}. In the second phase, the subjects performed another series of 2 coloring sessions in which the initial color of each node was taken from a $30$ second or $5$ second checkpoint of a session from the first phase. Thus, the subjects restarted the coloring task from a point in the middle of a previous session. However, they did not know from which session they were restarting, so they had no knowledge of the history of color changes that led to the checkpoint.

More precisely, from each session in {\em phase 1} that lasted more than $35$ seconds, we saved snapshots of the state of the session at $5$ seconds and at $30$ seconds. A snapshot is a mapping of participants to nodes and from nodes to current color at that point in time. In {\em phase 2}, the subjects were restarted from one of these snapshots. This means that subjects were mapped to the same nodes as in the snapshot, and nodes began with the same color as in the snapshot.

We used three different topologies (figure \ref{RestartTopos}) for the restart experiment, a simple cycle with no cords, a barbell network consisting of two cycles connected by a single edge, and a line graph. These graphs were selected to maximize the expected time to completion, thus reducing the expected number of session in the first phase to complete before $35$ seconds.
\begin{figure*}[t]
\begin{center}
  \includegraphics[width=0.55\textwidth]{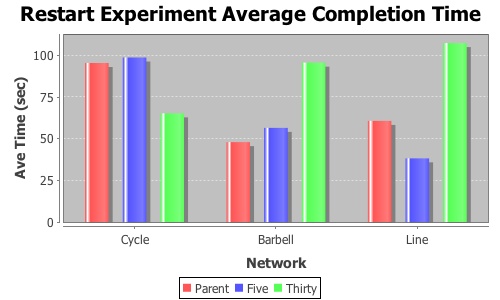}
  \caption{The average completion time of the {\em restart} experiments for each network topology.}
  \label{AveRst}
\end{center} 
\end{figure*}

\section{Analysis}

For both experiments, we are interested in detecting some form of history usage by the subjects and in measuring its positive or negative effects. If subjects do use significant portions of local history when making decisions, and if that use has a significant effect on performance, then it may be possible to detect this in completion time for the experiments. Figure \ref{AveSwp} gives the average completion time for the swap experiments of each topology for swaps of every $5$ seconds, $10$ seconds, and never.

This experiment has a number of drawbacks that result in a difficult analysis. Swapping erases the subjects local history knowledge, but it also may result in several unintended treatments including the confusion of switching nodes and the distribution of different subject strategies around the network during the course of the game. The execution of the experiment also suffered from a more practical drawback. Half of the $5$ and $10$ second swap experiments completed after only one or zero swaps, thus rendering the treatment relatively useless. A more elaborate analysis is needed to understand the experimental output.

We perform a similar analysis on the restart experiments. If we assume that there is no influence of history on performance, then the expected time to completion from a given checkpoint should be the same whether the subjects continue playing through the checkpoint time ({\em phase 1}) or whether they restart from that checkpoint later in the experiment ({\em phase 2}). Thus, if we can detect a significant difference in time to completion for the restarts, then may provide evidence of history usage.

However, in practice there may be an effect of restarting the participants from a checkpoint that has nothing to do with history. The subjects might require a context switching time to adapt to their new colors and those of their neighbors, or they might be influenced to choose colors more rapidly because of the restart. To correct for this, we restart the participants from a $5$-second and $30$-second checkpoint. Restarting from a $5$-second checkpoint creates the restart effect without erasing much history, while the $30$-second checkpoint has a restart effect and erases a significant amount of history.

It is interesting to visualize the dynamics of the sessions in the restart experiments. One way to visualize this is to plot the evolution of the Hamming distance to a correct coloring. The Hamming distance of a session at a particular point in time is computed in the following way. Pick an arbitrary 2 coloring solution. For each node in the network, if the node is consistent with that 2 coloring, assign it value $1$, if it has a color but is inconsistent, assign it value $-1$, and if it has no color, assign it value $0$. The Hamming distance of the session is the sum of these values for all nodes.

\begin{figure*}[!t]
  {\includegraphics[width=0.54\textwidth]{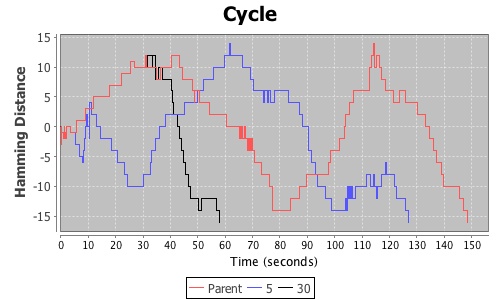}}
  \caption{Hamming distance versus time}
  \label{fig:hamm61}
  \end{figure*}

Figure \ref{fig:hamm61} describes the evolution of Hamming distance versus time for one session from {\em phase 1} along with the same plot for its $5$ second and $30$ second restarts from {\em phase 2}. This session was performed on the simple cycle with no cords. The red line gives the evolution of the Hamming distance for the {\em phase 1} session while the blue and black lines give that evolution for the 5 second and 30 second restarts respectively. Note the long periods of steady increase or decrease in Hamming distance for all three sessions represented. This is unlikely in a history-less setting and gives some of the intuition behind the directionality of conflicts described in section \ref{confbias}.

We lack a formal analysis of the dynamics of the restart experiment. In particular, we would like to formally capture any effects of the restart treatment on the dynamics of the sessions.

Figure \ref{AveRst} gives the average completion times for the three topologies used in the restart experiment. We suffer from a small dataset ($3$ or $4$ trials per topology) combined with a high variance in completion time. As the plots reveal we need a more finely tuned approach to detect any effect of history in the experiments, given the limited data.

The analysis described in section \ref{confbias} offers such an approach. Unfortunately, the conflict resolution bias does not apply to the restart experiments. The conflict resolution bias usage of historical information from the past few seconds while the restart experiments attempt to erase historical knowledge on longer time scales ($30$ seconds). However, the swapping experiments do offer an opportunity to apply the conflict resolution bias in a more interesting way. In the swapping experiments, subjects' historical knowledge is erased frequently. If participants do not use historical information, then the frequency with which they are swapped should have no effect on the conflict resolution bias. Thus, if the frequency of swapping does have an effect on the conflict resolution bias, then we have more evidence for the usage of short term history by the subjects.

In fact, there is a small change in the conflict resolution bias as the frequency of swapping varies. Table \ref{confbiasswap} gives the conflict resolution bias for each swap frequency used in the experiment. We note that the change in conflict resolution bias is small. This is again a reflection of the fact that conflict resolution bias only captures a limited time frame of history usage. We also note that the conflict resolution bias tends towards $0.5$ as the time between swaps approaches $0$. This is expected, as a shorter time between swaps gives subjects less of an opportunity of observing the timing of neighboring nodes' most recent color changes.

\begin{table}
\vspace{1em}
\centering
\begin{tabular}{|l|l|lr|}
\hline
\textbf{Time Between Swaps} & \textbf{Conflict Resolution Bias} \\
\hline
\hline
Never swap & 0.413 \\
\hline
10 seconds & 0.419 \\
\hline
5 seconds & 0.449 \\
\hline
\end{tabular}
\caption{Conflict resolution bias for each swap frequency.}
\label{confbiasswap}
\end{table}

\section{Discussion}

Our analysis focuses on crude aggregate measures (average completion time) or finely tuned measures that only capture short term history usage. It is conceivable that subjects are employing more elaborate, longer term usage of history. It may be possible to generalize conflict resolution bias to account for richer uses of history. In addition, it is possible that different human subjects employ different uses of history, and so it could be interesting to understand this distribution of strategies.

\vspace{1em}
\section{Acknowledgements}

The authors would like to thank Devin Barr for developing the experimental platform, as well as Daniel Enemark for useful discussion on the experimental design and for help conducting the experiments. This work is supported by NSF Grant No. 0905645.

\bibliography{CoordinationGames}

\end{document}